\documentclass[final, 5p]{elsarticle}
\usepackage{amsmath,amssymb}
\usepackage{lineno}
\modulolinenumbers[1]
\usepackage{booktabs}
\usepackage{threeparttable}
\usepackage{dcolumn}
\usepackage{epstopdf}
\usepackage[figuresright]{rotating}
\usepackage{xcolor}
\usepackage{multirow}
\usepackage{tabularx}
\usepackage{amsfonts}
\usepackage{textcomp}
\usepackage{wasysym}
\usepackage{graphicx}
\usepackage{bm}

\newcolumntype{b}{X}
\newcolumntype{s}{>{\hsize=.5\hsize}X}

\journal{Acta Materialia}

\biboptions{sort&compress}

\begin{document}

\begin{frontmatter}
	\title{Dissociated dislocation-mediated carbon transport and diffusion in austenitic iron} 
	
	\author[ampkth]{Ruiwen Xie}
	
	\author[ampkth]{Song Lu\corref{mycorrespondingauthor}}
	\cortext[mycorrespondingauthor]{Corresponding author}
	\ead{songlu@kth.se}
	\author[ampkth]{Wei Li}
	\author[nu1,nu2]{Yanzhong Tian}
	
	\author[ampkth,pauu,ispohu]{Levente Vitos}

	\address[ampkth]{Applied Materials Physics, Department of Materials Science and Engineering, Royal Institute of Technology, Stockholm SE-10044, Sweden}
	\address[nu1]{Key Laboratory for Anisotropy and Texture of Materials (Ministry of Education), School of Materials Science and Engineering, Northeastern University, Shenyang 110819, China}
	\address[nu2]{State of key laboratory of rolling and automation, Northeastern University, Shenyang 110819, China}
	\address[pauu]{Department of Physics and Astronomy, Division of Materials Theory, Uppsala University, Uppsala, Sweden}
	\address[ispohu]{Wigner Research Centre for Physics,Institute for Solid State Physics and Optics, Budapest, Hungary}
	
\begin{abstract}
	Dislocation-solute interaction plays fundamental roles in mechanical properties of alloys. Here, we disclose the essential features of dislocation-carbon interaction in austenitic Fe at the atomistic scale. We show that passage of a Shockley partial dislocation in face-centered cubic iron is able to move carbon atoms on the slip plane forward by one Burgers vector, revealing a novel dissociated dislocation-mediated transport mechanism. This mechanism is induced by shear, which is distinct from the normal thermally activated diffusion process. Furthermore, we show that there exists a fast diffusion channel with significantly reduced diffusion energy barrier in the partial dislocation core, which is highly localized and directional. These inherent geometrical features are crucial for understanding the dependence of the diffusivity of dislocation pipe diffusion on the character of dislocations; most importantly, they can result in unbalanced pinning effect on the leading and trailing partials in a mixed dislocation, consequently facilitating stacking fault formation and deformation twinning. This explains the controversial effects of carbon on deformation twinning observed in various alloys.  Our findings pave the road to tune mechanical properties of materials by manipulating dislocation-interstitial interaction.
\end{abstract}
	
\begin{keyword}
	\texttt Carbon-dislocation interaction \sep Dislocation-mediated carbon transport \sep Fast diffusion channel \sep Unbalanced pinning-aided twinning
\end{keyword}
	
\end{frontmatter}








\section{\label{sec:intro}Introduction}

Carbon is undoubtedly one of the most important alloying elements in metallurgy, especially in steels. Thermodynamically, it is well accepted that C is a strong austenite stabilizer and noticeably increases the stacking fault energy (SFE) \cite{LU201772}, consequently C addition should suppress deformation twinning (DT) in austenite according to the classical plasticity theory \cite{DECOOMAN2018283}. Controversially, it is often observed that C addition strongly promotes DT in various face-centered cubic (fcc) alloys including the important engineering alloys such as high-Mn twinning-induced plasticity (TWIP) steels \cite{KOYAMA201720,Lee3573,BOUAZIZ2011141,koyama2018overview}, Fe-Ni-C austenitic steels \cite{SACHDEV198731} and the novel high/medium-entropy alloys like Fe-(30-40)Mn-10Co-10Cr (at.\%) \cite{li2017interstitial,CHEN2018150},  CrMnFeCoNi \cite{WU2015815,LI2019400} and CrCoNi \cite{SHANG201977}. The C-enhanced twinnability in these alloys is a crucial factor for the attainment of the high work-hardening capacity and the excellent mechanical properties such as a synergy of high strength and high ductility \cite{BOUAZIZ2011141}. Although the exact microscopic mechanism remains elusive, the prominent effect of C on DT is usually ascribed to dislocation-C interaction \cite{BOUAZIZ2011141,koyama2018overview} which also underlies other fundamental strengthening mechanisms such as solid solution hardening~\cite{Cottrell}, dislocation planar slip~\cite{DECOOMAN2018283,HONG19901581,LEE20116809} and dynamic strain aging (DSA)~\cite{DECOOMAN2018283,koyama2018overview}. Today, the microscopic features and mechanisms of dislocation-C interaction, especially at the dislocation core, are generally unclear and not accessible by experiments~\cite{RODNEY2017633}.  The lack of essential understanding of dislocation-C interaction has caused spirited disputations regarding the influence of C addition on the essential factors like the SFE, DT, DSA as well as martensitic phase transformation (MT) that strongly affect the plastic properties \cite{DECOOMAN2018283,BOUAZIZ2011141,koyama2018overview}.  

Dislocation is usually believed to significantly accelerate solute diffusion along the dislocation line, referred to as the dislocation pipe diffusion (DPD), which is ascribed to the reduced activation barrier \cite{LOVE1964731,Balluffi,Huang1989}. The low activation barrier of DPD,  normally assumed 40\%-80\% of the corresponding bulk diffusion values for substitutional solutes and less than $\sim$50\% for interstitial elements \cite{Balluffi,Huang1989,PICU2004161,Legros1646}, presumably originates from the distorted atomic structure or the presence of vacancies at the dislocation core \cite{Huang1989}. However, recent studies elucidate that whether DPD can be faster than the normal bulk diffusion  depends remarkably on the characteristics of the dislocation \cite{ishii2013conjugate}. For example, screw dislocations in $\alpha$-Fe were even shown to slow down hydrogen diffusion \cite{Kimizuka2011}. Similarly, atomistic simulations demonstrated that solute diffusion along the grain boundaries (GBs) can be either accelerated or inhibited, governed by the local GB structures at the atomic level \cite{zhou2019}.

In order to understand the influence of C on the aforementioned phenomena and processes, it is of significance to reveal the essential structural features that control C migration behavior at/near the dislocation core. In fcc materials with low SFEs, the $ a $/2 $ < $110$ > $ full dislocation normally dissociates into two $ a $/6 $ < $112$ > $ Shockley partials bounding a stacking fault (SF). Here, using quantum-mechanical ab initio calculations, we explore the migration behaviors of C in bulk, SF and partial dislocation core in the double-layer antiferromagnetic (AFMD) $\gamma$-Fe ($Supplementary~Information, SI$, Fig.S1) \cite{Boukhvalov2007} with a focus of understanding the influence of C on dislocation mobility in high-Mn steels which have N\'{e}el temperatures at around room temperature \cite{ALLAIN2010500,endoh1971antiferromagnetism}. We observe two striking consequences of the partial dislocation-C interaction, both of which arise from the fact that the Shockley partial dislocation exchanges the characteristics of the octahedral ($ O $) and tetrahedral ($ T $) interstitial sites after its passage. First, passage of a Shockley partial dislocation transfers the C atoms on the slip plane forward by a Burgers vector, which gives rise to a unprecedented dissociated dislocation-mediated mechanism for C transport. Second, the fast diffusion channel with significantly reduced energy barrier at the partial dislocation core is strongly localized and highly directional, i.e., perpendicular to the Burgers vector of the partial dislocation, which unveils the atomistic origin of the dependence of DPD on the characteristics of the dislocation. The above findings are crucial for understanding the observed C effects on dislocation planar slip, DSA, DT and MT in a consistent picture, which are beyond the thermodynamic role played by C.

\section{Computational Methodology}
The equilibrium structure of AFMD $\gamma$-Fe with alternate two-layer collinear up- and down- spins along the [001] direction is fully relaxed. The obtained lattice and magnetic parameters  are summarized in $SI$ Table S1, in comparison with available data in literature. The AFMD $\gamma$-Fe is stabilized in the face-centered tetragonal structure, however, here it is still denoted as fcc for simplicity. The tetragonality is considered for all the calculations. The MES calculations are performed using a supercell with six (111)  layers along the \textbf{\emph{c}} direction. The \textbf{\emph{a}} and \textbf{\emph{b}} lattice vectors of the supercell are 3$ \times $$a$/2[01\={1}] and $a$[\={2}11], respectively (Fig.~\ref{scheme} (A)). The GSF structures are created by tilting \textbf{\emph{c}} axis along the [11\={2}] direction by the shear vector \textbf{\emph{u}}, i.e., \textbf{\emph{c}} = \textbf{\emph{c}} + \textbf{\emph{u}} \cite{kibey2006effect}. One C atom is placed on the slip plane. The nominal concentration of C is $\sim$1.4 at.\% ($\sim$0.3 wt.\%), and the C-C interaction due to the periodic boundary conditions on the results is negligible. In order to obtain the MES, we calculate the total energy of the supercell by moving C step by step in the rectangular areas on the slip plane (Fig.~\ref{scheme} (A)). The MEP is then identified on the MES, which is a straight path along the $<$110$>$ direction. This is consistent with the previous result obtained by using the nudged elastic band method \cite{Jiang2003,hepburn2013first}. For the MES calculations, atomic relaxation to the second-nearest coordination shells of C is performed. With the current setup for modeling dislocation core/near-core structures, full atomic relaxation is not allowed. The effect of this constrained relaxation on the diffusion energy barriers is discussed in $SI$ text, which does not affect our findings. Similar calculations are performed in the nonmagnetic (NM) and single-layer antiferromagnetic (AFMI) $\gamma$-Fe, as well as ferromagnetic (FM) Ni, the obtained results are consistent with those reported here and presented in $SI$. 

All the calculations are carried out using the Vienna ab-initio simulation package (VASP)~\cite{kresse1993ab,kresse1996efficient}. The electron-ion interactions are described with the projector-augmented-wave method (PAW)~\cite{kresse1999ultrasoft,blochl1994projector}. The Perdew-Burke-Ernzerhof formula for the generalized gradient approximation is adopted for the exchange-correlation functional~\cite{perdew1996generalized}. The energy cut-off for the plane-wave basis sets is 350 eV. The k-point mesh of 4$\times$4$\times$2 for the supercell calculations is generated using the Monkhorst-Pack scheme~\cite{monkhorst1976special}. The forces on atoms are converged to less than 0.02 eV \AA$^{-1}$ when atomic relaxation is allowed. 

\section{Results and Discussion}
\subsection{Migration Energy Surface}

\begin{figure*}[t!]
	\centering
	\includegraphics[width=1\linewidth]{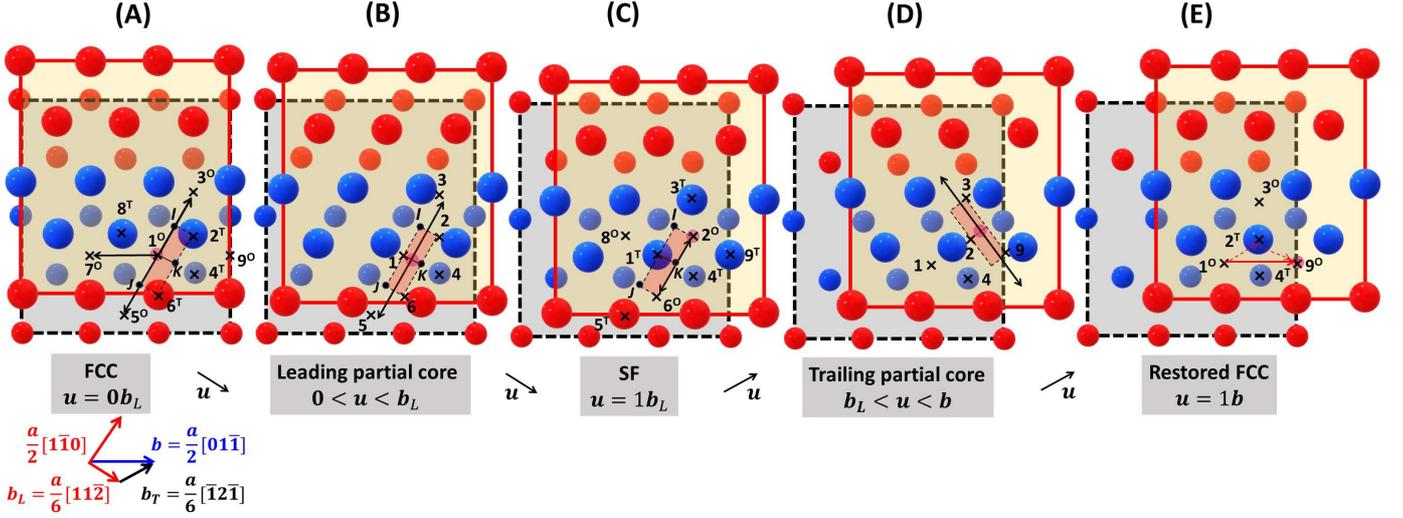}
		\caption{Schematics of local atomic structure variation around a C atom during the passage of a pair of leading and trailing partial dislocations. The red and blue colors indicate the up- and down-spins, respectively, in the AFMD $\gamma$-Fe. The C atom is placed at the interstitial site between two (111) slip planes (small and large atoms, respectively). (A-E) The local structures for the original fcc ($\bm{u}=0\bm{b_{L}}$, $\bm{b_{L}}=a/6[11\overline{2}]$), leading partial core/near-core ($0<\bm{u}<\bm{b_{L}}$), stacking fault ($\bm{u}=1\bm{b_{L}}$), trailing partial core/near-core ($\bm{b_{L}}<\bm{u}<\bm{b}$, $\bm{b}=a/2[01\overline{1}]$) and restored fcc ($\bm{u}=1\bm{b}$), respectively. Even and odd numbers denote the positions of the interstitial $O$ and $T$ sites, respectively, found on the slip plane in the fcc lattice. $I$ and $J$ denote the positions of crowdion sites on the $<$110$>$ diffusion paths, $1^O_{fcc}\rightarrow3^O_{fcc}$ and $1^O_{fcc}\rightarrow5^O_{fcc}$, respectively, in the original fcc lattice. The migration energy surface is studied in the shaded rectangle areas ($1K2I$ and $1K6J$).}
	    \label{scheme}
\end{figure*}

When a leading partial dislocation with Burges vector \textbf{\emph{b$_{L}$}} glides towards a C atom on the close-packed \{111\} plane, the local structural environment around the C atom evolves correspondingly. Specifically, the leading partial creates a SF after its passage. Here, we use the shear vector \textbf{\emph{u}} to measure the structure evolution from fcc (\textbf{\emph{u}} = 0\textbf{\emph {b$_L $}}) to SF (\textbf{\emph{u}} = 1\textbf{\emph {b$_L $}}) in a continuous way (Fig.~\ref{scheme}). For 0\textbf{\emph {b$_L $}} $ < $ \textbf{\emph{u}} $ < $ 1\textbf{\emph {b$_L $}}, it corresponds to the generalized stacking fault (GSF), which models the partial dislocation core or near-core structures. It is important to observe that the leading partial transforms all the interstitial $O$ sites on the slip plane in fcc lattice to the $T$ sites in the SF (e.g., $1_{fcc}^O\rightarrow1_{SF}^T$) and only half numbers of the neighboring $T$ sites in fcc lattice to the $O$ sites in the SF (e.g., $2_{fcc}^T\rightarrow2_{SF}^O$), whereas the rest $T$ sites maintain their character (e.g., $4_{fcc/SF}^T$, Fig.~\ref{scheme}). Since C atoms prefer $O$ sites in both fcc lattice and the SF \cite{gholizadeh2013influence}, passage of a partial dislocation is expected to provoke the movement of the C atoms on the slip plane. The emerging question to address is when the process should happen, during or after the passage of the partial dislocation, which strongly affects how C atoms impede the mobility of both leading and trailing partials.

\begin{figure*}[t!]
	\centering
	\includegraphics[width=1.05\linewidth]{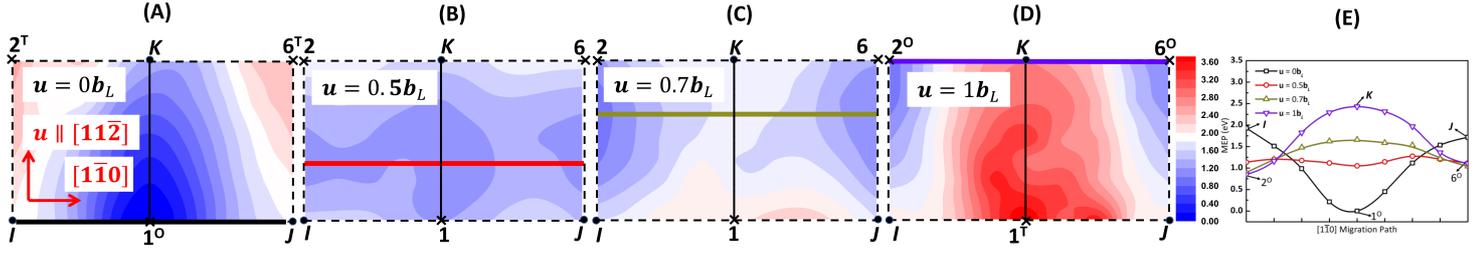}
		\caption{Migration energy surface (MES) on the (111) slip plane.  (A-D) The MES in the $1K2I$ and $1K6J$ rectangle areas in fcc  ($\bm{u}=0\bm{b_L}$), dislocation core/near-core ($\bm{u}=0.5\bm{b_L}$ and $\bm{u}=0.7\bm{b_L}$) and stacking fault ($\bm{u}=1\bm{b_L}$), respectively. All the energies are relative to the total energy of the fcc structure with the C atom at the 1$^O_{fcc}$ site.  (E) The minimum migration energy path (MEP) for various $\bm{u}$, as marked on the MES in (A-D) by the thick solid lines.}
	\label{MEP}
\end{figure*}

In order to probe the migration behavior of C as the local structure evolves from fcc to SF, we calculate the relevant migration energy surface (MES, Fig.~\ref{MEP}) in the rectangle areas $1K2I$ and $1K6J$ in Fig.~\ref{scheme} (A-C), with respect to \textbf{\emph{u}}. Here, \textbf{\emph{u}} is parallel to the $a$/6[11$\overline{2}$] direction on the (111) slip plane, which has a significantly lower slip barrier than slipping along the $a$/6[$\overline{2}$11] direction due to the magnetic ordering in AFMD $\gamma$-Fe ($SI$, Fig.S1 (E)). For the same reason, the MES in the $1K2I$ and $1K6J$ areas are not symmetric (Fig.~\ref{MEP} (A-D)). In bulk (\textbf{\emph{u}} = 0\textbf{\emph {b$_L $}}), the C atom occupies site 1$^O$ as indicated by the energy minimum in Fig.~\ref{MEP} (A). The two neighboring $T$ sites, 2$_{fcc}^T$ and 6$_{fcc}^T$, are unstable for C. As the leading partial moving close to the C atom (increasing \textbf{\emph{u}}), the stable position for C gradually moves away from site 1 along the shear direction and changes to a position nearby site 2 (or 6) (Fig.~\ref{MEP} (B-C)). Notice that all the interstitial sites gradually lose the geometrical characters of $O$ or $T$ sites when \textbf{\emph{u}} approaches 0.5\textbf{\emph {b$_L $}} (Fig.~\ref{MEP} (B)). For \textbf{\emph{u}} = 1\textbf{\emph {b$_L $}}, interstitial sites 2 and 6 at the SF become the $O$ sites, being the energetically favorable positions for C (Fig.~\ref{MEP} (D)). It is important to observe that once \textbf{\emph{u}} is larger than a critical value, $ \sim $0.7\textbf{\emph {b$_L $}}  (Fig.~\ref{MEP} (C)), there is no energy barrier on the MES hindering the C atom moving from the vicinity of the original site 1 to the more stable positions near site 2 or 6, as the partial dislocation moves forward. 

\subsection{Dissociated Dislocation-mediated Transport of C}
The above results clearly show that when a leading partial dislocation approaches a C atom on the slip plane, changing its surrounding environment from fcc to the GSF-like structures, the C atom will automatically adapt its position and shuffles from the original $O$ site in fcc to the more stable position at the GSF. This movement of C is incorporated into the movement of the Fe atoms in the partial core upon external stress, therefore should occur instantly, that is, as fast as the speed of the dislocation. Notice that in the present magnetic configuration, a C atom has two potential final positions to choose, sites 2 and 6 (Fig.~\ref{MEP} (D)), which differ slightly in energy due to the local magnetic environment; and the final choice in metals and alloys should be affected by the fine dislocation core structures, local magnetic moments  as well as chemistry \cite{SMITH2016352}.  Nevertheless, we can conclude that there is no thermodynamic or kinetic reason for C atoms to be transferred to (or left at) the unfavorable $T$ sites (4$^T_{SF}$ or 1$^T_{SF}$ in Fig.~\ref{scheme} (C)) after the passage of a partial dislocation. Therefore, the present finding revokes all the previous arguments/mechanisms constructed on the assumption that the leading (/twinning) partials transfer C atoms to the unstable $T$ sites in the SF (/twin) area \cite{Adler1986,LEE20116809,JUNG20136724,DECOOMAN2018283}. For example, Adler et al.~\cite{Adler1986} ascribed the exceptionally large work hardening capacity in Hadfield steels to the tetragonal distortion caused by C atoms at the $T$ sites in deformation twins. By contrast, Lee et al. \cite{LEE20116809} argued that C atoms at the $T$ sites at the SF area can easily hop to the $O$ sites below the slip plane in Fe-Mn-C steels, which leads to the reorientation of Mn-C pairs and pinning of the trailing partial. Unfortunately, such unfounded microscopic picture has been widely adopted to explain the occurrence of DSA in high-Mn TWIP steels at room and lower temperatures when C bulk diffusion is too slow to account for the pinning process underlying DSA ~\cite{LEE20116809,JIN20121680,JUNG20136724,DECOOMAN2018283,koyama2018overview,YANG2017146,YANG2014551}.

Following the same interaction mechanism between the leading partial and C atoms, when a trailing partial restores the local stacking sequence from SF to fcc (Fig.~\ref{scheme} (C-E)), again it  pushes C atoms in the SF area forward on the slip plane by one step, e.g., 2 $\rightarrow$ 9 in Fig.~\ref{scheme} (E). Therefore, the overall effect of passing a pair of partials is to displace the interstitial C atoms  on the slip plane by a Burgers vector of $ a $/2$<$01\={1}$>$, e.g., 1 $\rightarrow$ 9 in Fig.~\ref{scheme} (E), realizing the novel dissociated dislocation-mediated transport mechanism. In literature, it is generally accepted that at low strain rates, slowly moving dislocations can drag a cloud of solute atoms via bulk diffusion (Cottrell atmosphere~\cite{Cottrell}), whereas at high strain rates and low temperatures, the bulk diffusivity of solutes is too slow to catch up with the rapidly moving dislocations. Hence, the solutes are thought to be stationary~\cite{ESTRIN19862455}. Here, we reveal that in low-SFE materials, besides the thermally activated diffusion, moving dissociated dislocations provide an inherent shear-induced mechanism for the local transport of C atoms. This observation is in fact consistent with the nature of dislocations, that is to spatially transport mass as characterized by the Burgers vector. Considering the enormous numbers of dislocations during deformation, the currently disclosed mechanism can lead to significant redistribution of C atoms, especially when the normal diffusion hops are strongly suppressed at cryogenic temperatures. Furthermore, we envisage that successive passages of dissociated dislocations will  gradually sweep C atoms away from dislocation sources like Frank-Read sources or GBs, which creates a C depletion area in the front of the source and generates a locally softened glide plane with low SFE and low frictional stress for subsequent dislocations. Consequently, the planarity of dislocation slip is amplified by the positive feedback due to this local C transport mechanism at the atomistic scale. The present finding offers a new perspective for understanding the effect of interstitials (C, N) on promoting planar slip in austenitic steels~\cite{GUTIERREZURRUTIA20116449,KOCKS1985623}, which is distinct from the extant theories~\cite{OWEN1998111,DECOOMAN2018283,GEROLD19892177}.

\subsection{Fast Diffusion Channel}
Previous studies showed that the minimum migration energy path (MEP) for C diffusion in bulk $\gamma$-Fe is a direct path along the $<$011$>$ direction connecting two neighboring $O$ sites with the transition state located at the crowdion site~\cite{Jiang2003,hepburn2013first}, which is also confirmed in Fig.~\ref{MEP} (A). The obtained diffusion energy barriers along various $<$011$>$ paths with different local magnetic configurations are consistent with the previous calculations \cite{hepburn2013first} (for detailed comparison, see $SI$, Table S1). For \textbf{\emph{u}} $ > $ 0\textbf{\emph {b$_L $}}, the MEP is shifted along the slip direction, but maintains a nearly straight path (Fig.~\ref{MEP} (B-D)). Remarkably, in the representative partial core/near-core structures (Fig.~\ref{MEP} (B) and (C)), a fast diffusion channel with significantly reduced diffusion energy barriers is opened up. In particular, the diffusion energy barrier along the MEP for \textbf{\emph{u}} = 0.5\textbf{\emph {b$_L $}} is $\sim$0.22 eV, corresponding to $\sim$13\% of that for the same path in bulk (1.72 eV, Fig.~\ref{MEP} (E)). The formation of the fast diffusion channel mainly has a geometrical origin, which is verified by calculations at other magnetic states (see details in $SI$). An unstaggered avenue along the $\pm$[\={1}10] directions is formed upon shearing along the Burgers vector of $a$/6[11\={2}] at \textbf{\emph{u}} $ \approx $ 0.5\textbf{\emph {b$_L $}}, which locally removes the geometrical characteristics of the $O$ and $T$ sites and flattens the migration energy landscape. It provides a channel with large free volume which allows the interstitial C atoms to move easily \cite{ishii2013conjugate}. The formation of such unstaggered avenue at the partial core was indeed observed in the previous molecular dynamics simulations \cite{ishii2013conjugate,SMITH2016352,TSUZUKI20091843}. We mention that in the real case, for example, when an edge partial dislocation generates asymmetrical compression and tension strain fields above and below the slip plane, one should not expect the perfect unstaggered channels in the core or near-core regions. 

\begin{figure}[t!]
	\centering
	\includegraphics[width=\linewidth]{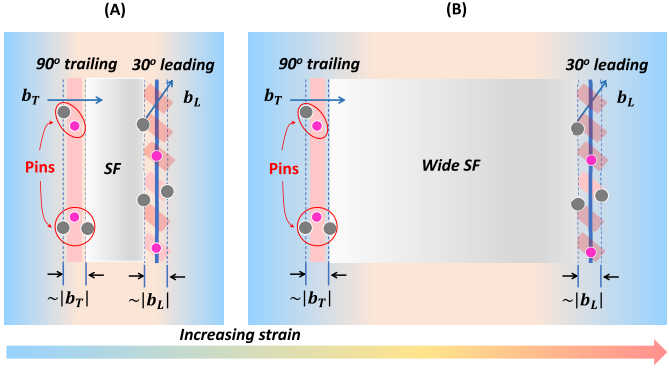}
		\caption{Schematic for unbalanced pinning effect in a dissociated 60$ ^{\rm o} $ dislocation. (A) The fully interconnected fast diffusion channel is formed along the 90$^{\rm o}$ trailing partial line, which facilitates the formation of local pins including magnetic and chemical SROs via C fast redistribution in the core; whereas in the core of the 30$^{\rm o}$ leading partial, the fast diffusion channels are not well connected to each other, thereby C diffusion is retarded and less pins can be formed. (B) A wide SF is formed between the two partials upon shearing, assisted by the unbalanced pinning effect.}
	\label{pinning}
\end{figure}

Importantly, the present work allows us to identify the critical geometrical features of the fast diffusion channel. From Fig.~\ref{MEP} (B-C), we observe that the fast diffusion channel at/near the partial dislocation core is highly directional, i.e., perpendicular to the Burgers vector of the partial dislocation. Inherently, the length of this channel is confined by the partial core width, $\sim \vert$\textbf{\emph {b$_L $}}$\vert$ \cite{zhou2019}; in other words, it is strongly localized in nature. Only when a partial is of pure edge character (90$^{\rm o}$), the localized diffusion channels are interconnected and form a long fast diffusion path along the dislocation line, otherwise, misalignment is expected between the localized fast diffusion channels and the dislocation line (Fig.~\ref{pinning} (A)). Consequently, the actual diffusivity of DPD along a single dislocation depends strongly on the dislocation character~\cite{PICU2004161}, and the net diffusion along a mixed dislocation line generally requires the aid of out-core diffusion~\cite{Zhang012101,Heuser2014}, or of the local thermally activated vibrations of dislocation line itself~\cite{TAPASA200793}. We may mention that the present finding is akin to the recently disclosed roles of GBs in affecting hydrogen diffusion, showing that only high-angle GBs are able to facilitate hydrogen diffusion via forming interconnected low-barrier channels  \cite{zhou2019}.

Interestingly, the SF ribbon itself also possesses lower in-plane diffusion barriers compared to those in bulk (Fig.~\ref{MEP} (E) and $SI$ Table S1), which is consistent with the previous theoretical and experimental results regarding the roles of SF in DPD in dissociated dislocations \cite{Huang1989,Zhang012101,Tang2012}. Considering that the outward-diffusion of C atoms from the SF plane as driven by its effect on increasing the SFE is very slow at room temperature (``anti-suzuki effect''~\cite{HICKEL2014147}), the fast redistribution of C at the SF is considered to be mostly confined between the two slip planes, similarly as self-interstitials and vacancies \cite{Huang1989,PICU2004161}.  

\subsection{Unbalanced Pinning}
A significant consequence of presently disclosed geometric relationship between the fast diffusion channel and the Burgers vector of a partial dislocation is the unbalanced pinning effect (Fig.~\ref{pinning}). In general, C impedes the movement of dislocations by introducing local lattice distortion and increasing the overall lattice frictional stress. When randomly distributed C atoms are incorporated into the dislocation core and redistribute to lower the total energy of the dislocation-C system, driven by magnetism or chemistry induced fluctuations on the migration energy surface along the dislocation line, the dislocation is subject to an excess pinning force. The pinning force therefore depends on whether C atoms can effectively move to seek for the low-energy positions. When this process is sufficiently realized on the temporarily arrested dislocations, it can cause DSA \cite{Koyama2011342,Dastur1981,koyama2018overview}. We note that the bulk diffusivity of C in austenitic steels is extremely slow at room temperature due to the high activation barriers (see $SI$ Table S1), which can not account for the pronounced DSA effect as macroscopically manifested by the serrations on the stress-strain curves at room and cryogenic temperatures \cite{koyama2018overview}; whereas the redistribution within the localized fast diffusion channel can be easily accomplished as indicated by the very low migration energy barriers (Fig.~\ref{MEP} (E)). This excess pinning force is therefore maximized on the pure edge partial with fully interconnected fast diffusion channels along the dislocation line which efficiently facilitates C redistribution; but minimized on the pure screw partial with short and separated diffusion channels. As a consequence, the unbalanced (unequal) pinning is generally expected on the leading and trailing partials in a dissociated dislocation of mixed character (Fig.~\ref{pinning}). 

\begin{figure}[t!]
	\centering
	\includegraphics[width=\linewidth]{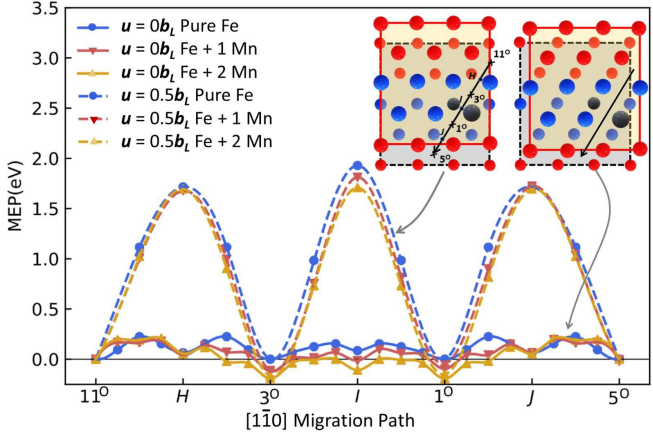}
		\caption{Manganese-induced rugged migration energy landscape. Dashed and solid lines are the MEPs in bulk and partial core ($\bm{u}=0.5\bm{b_L}$), respectively. Manganese atoms (gray balls) generate low energy sites and promotes C redistribution in the partial core, causing pinning.}
	\label{mnmep}
\end{figure}

In AFMD $\gamma$-Fe, pure magnetism-induced energy variation between different potential (meta)stable positions for C in the core is small ($\sim$0.1 eV, Fig.~\ref{MEP} (E)).  However, in alloys one can expect a remarkably uneven energy landscape with abundant attractive low-energy sites available in the core at finite temperatures, which is likely to be further enhanced when chemical and magnetic short range order (SRO) or fluctuations are met \cite{VonAppen2011,Ikeda2019}. In Fig.~\ref{mnmep}, we demonstrate that even one or two Mn atoms can generate notably deeper pinning sites in the core, which will attract C atoms to form Mn-C SROs and obstruct the mobility of the partial. This is consistent with the previous ab initio studies showing that Mn and C atoms tend to form SROs in bulk $\gamma$-Fe \cite{VonAppen2011}. Thereby, dislocations or segments with stronger pinning force on the trailing partials than that on the leading ones can be more easily separated into wide SF ribbons under applied stress, which consequently facilitates the nucleation of deformation twins or hexagonal close-packed (hcp, $\varepsilon$) martensites via the SF overlapping mechanisms (Fig.~\ref{pinning} (B))~\cite{GUTIERREZURRUTIA20116449,BYUN20033063,koyama2018overview}. Such unbalanced pinning-aided mechanism for DT is essential for understanding the observation that C addition notably promotes DT in various alloys, despite the fact that C increases the SFE \cite{SACHDEV198731,BOUAZIZ2011141,koyama2018overview,WU2015815,LI2019400,CHEN2018150,Ikeda2019}. The critical role of unbalanced pinning on DT is clearly demonstrated by the distinct deformation mechanisms in Fe-33Mn (wt.\%, SFE = 37 mJ m$^{-2}$) and Fe-33Mn-1.1C (wt.\%, SFE = 57 mJ m$^{-2}$) which are dislocation glide and DT, respectively \cite{KOYAMA201720}. Similar arguments apply for the deformation-induced $\gamma-\varepsilon$ MT, where C was observed to accelerate the kinetics of the phase transformation \cite{SEOL2013558,SEOL2017187}.  We note that C addition causes solid solution hardening, increasing the flow stress, and thus the critical twinning stress may be reached at smaller strains in C-contained alloys. However, this effect is usually weak for interstitial element in austenite \cite{BOUAZIZ2011141}; further, recent studies show that only $\sim$ 0.2-0.8 at.\% C addition can effectively promote DT \cite{WU2015815,LI2019400,SHANG201977}, which obviously cannot be ascribed to the negligible solid solution hardening effect.

Furthermore, since unbalanced pinning originates from C diffusion over the waiting time $t_w$ when dislocations are temporarily arrested at obstacles, its strength should negatively depend on strain rate ($\dot{\varepsilon}$). Here, $t_w$ is inversely proportional to $\dot{\varepsilon}$ \cite{Curtin2006}. Therefore, increasing stain rate will suppress DT/ MT via weakening the unbalanced pinning-aided mechanism, which is fully consistent with recent experimental observations in C-alloyed austenites~\cite{SEOL2013558,SEOL2017187,YANG2017146,YANG2014551,koyama2018overview}. We emphasize that the negative strain rate dependence of twinning activities in C-contained alloys is in stark contrast to the general effect of high strain rate which notably promotes DT in normal fcc metals or alloys like Al, stainless steels and C-free high-Mn steels~\cite{SEOL2017187,YANG2017146}. On the other hand, it is consistent with the negative strain rate sensitivity of flow stress through the suppression of DT, thereby weakening the dynamic Hall-Petch effect \cite{koyama2018overview,BOUAZIZ2011141}. Such unusual strain rate dependence therefore provides a strong evidence of our proposed unbalanced pinning picture. 

\section{Conclusions}

We studied the interaction mechanisms between partial dislocations and interstitial C atoms in $\gamma$-Fe and Fe-Mn steels at the atomistic scale. We discovered a so far unknown dissociated dislocation-mediated mechanism for C transport, which is crucial to account for the deformation-induced C redistribution, especially at low temperatures. This mechanism advances the current knowledge regarding the transport mechanism of interstitial elements (C, N), in addition to the normal diffusion process.  Furthermore, we disclosed the fundamental features of the fast diffusion channel at the partial dislocation core. The highly localized and orientated nature of the fast diffusion channel results in the general unbalanced pinning effect on the dissociated dislocations in C-contained alloys, which assists SF and twinning formation and competes with the effect of C on increasing the SFE. This mechanism is not limited to $\gamma$-Fe or Fe-Mn steels, but can be generalized to other advanced alloys with excellent mechanical performance, such as the C-alloyed medium and high entropy alloys \cite{li2017interstitial,CHEN2018150,WU2015815,LI2019400,SHANG201977,Ikeda2019}, for which the observations are also in perfect agreement with our theory. 
The present findings provide a physics-based understanding for the multiple roles played by C on DSA, dislocation planar slip and DT in various fcc alloys in a consistent picture and shed light on designing new alloys with interstitial-induced strengthening mechanisms.

\section*{Acknowledgments}
The authors thank Prof. Se Kyun Kwon and Prof. Han Soo Kim from POSTECH for valuable discussion. S.L., R.X., W.L. and L.V. thank the Swedish Research Council, the Swedish Foundation for Strategic Research, the Carl Tryggers Foundations, the Swedish Foundation for International Cooperation in Research and Higher Education, the Hungarian Scientific Research Fund (OTKA 128229), and the China Scholarship Council for financial supports. S.L. and L.V. are grateful for the financial support under the project ``Future Materials Design'' by the Swedish steel producers' association (Jernkontoret) and the Sweden's innovation agency (Vinnova). The Swedish National Infrastructure for Computing (SNIC) at Link\"oping is acknowledged for computation resource.

\end{document}